\documentclass[twocolumn,prl,superscriptaddress,showpacs,preprintnumbers,amsmath,amssymb,nofootinbib]{revtex4}

\usepackage{subfigure}
\usepackage{amsmath,amssymb}
\usepackage{graphicx}
\usepackage{rotating}
\usepackage{bm}
\usepackage{color}

\def\di{\displaystyle}

\def\bg{\begin{eqnarray}\begin{array}{rcl}\displaystyle}
\def\eg{\end{array} &\di    &\di   \end{eqnarray}}
\def\bm#1{\begin{eqnarray}\begin{array}{#1}\di}
\def\bmo#1{\begin{eqnarray*}\begin{array}{#1}\di}
\def\bml#1#2{\begin{eqnarray}\begin{array}{#1}\label{#2}\di}
\def\bgo{\begin{eqnarray*}\begin{array}{rcl}\displaystyle}
\def\ego{\end{array} &\di    &\di \nonumber  \end{eqnarray*}}

\def\btensor#1#2{\renew\left#1\begin{array}{#2}\di}
\def\brtensor#1#2#3{\ren#3\left#1\begin{array}{#2}}
\def\botensor#1#2{\renew\left#1\begin{array}{#2}}
\def\etensor#1{\end{array}\right#1}

\def\eq#1{(\ref{#1})}
\def\Eq#1{Eq.~(\ref{#1})}

\def\id{1\!\mbox{l}}

\def\s0#1#2{\mbox{\small{$ \frac{#1}{#2} $}}}
\def\0#1#2{\frac{#1}{#2}}

\def\CO{{\mathcal O}}
\def\CP{{\mathcal P}}

\def\ren#1{\renewcommand{\arraystretch}{#1}}

\def\renew{\renewcommand{\arraystretch}{1}}

\newcommand{\Tr}{\mathrm{Tr}}

\newcommand{\tr}{\mathrm{tr}}

\newcommand{\I}{\mathrm{i}}
\newcommand{\be}{\begin{eqnarray}}
\newcommand{\ee}{\end{eqnarray}}

\newcommand{\Gk}{\Gamma_k}

\newcommand{\Tc}{T_{\text{c}}}

\newcommand{\Nc}{N_{\text{c}}}
\newcommand{\lqcd}{\Lambda_{\text{QCD}}}
\newcommand{\pat}{\partial_t}
\newcommand{\Eqref}[1]{Eq.~\eqref{#1}}

\newcommand{\nul}{\nu_\ell}

\begin{document}

\title{Quark Confinement from Color Confinement}

\preprint{HD-THEP-07-22}


\author{Jens Braun}
\affiliation{TRIUMF, 4004 Wesbrook Mall, Vancouver, BC, Canada, V6T 2A3}
\author{Holger Gies}
\affiliation{Institut f\"ur Theoretische Physik, Universit\"at Heidelberg, Philosophenweg 16,
D-69120 Heidelberg, Germany}
\author{Jan M.~Pawlowski}
\affiliation{Institut f\"ur Theoretische Physik, Universit\"at Heidelberg, Philosophenweg 16,
D-69120 Heidelberg, Germany}

\begin{abstract}
  We relate quark confinement, as measured by the Polyakov-loop order
  parameter, to color confinement, as described by the
  Kugo-Ojima/Gribov-Zwanziger scenario. We identify a simple criterion
  for quark confinement based on the IR behaviour of ghost and gluon
  propagators, and compute the order-parameter potential from the
  knowledge of Landau-gauge correlation functions with the aid of the
  functional RG.  Our approach predicts the deconfinement transition
  in quenched QCD to be of first order for SU(3) and second order for
  SU(2) -- in agreement with general expectations. As an estimate for
  the critical temperature, we obtain {$\Tc\simeq 284\text{MeV}$} for
  SU(3).
\end{abstract}

\maketitle

\section{Introduction} 

Aside of the confinement of quarks, the confinement of gluons is a
challenging and unsolved problem. Various scenarios predict the
confinement mechanism to be manifest in the infrared domain of
gauge-dependent correlation functions. In the Kugo-Ojima
\cite{Kugo:1979gm} and Gribov-Zwanziger scenarios \cite{Gribov:1977wm}
(KOGZ) an infrared enhancement of the ghost and an infrared
suppression of the gluon signal confinement. These scenarios have been
investigated by a variety of non-perturbative field theoretical tools
such as functional methods
\cite{vonSmekal1997,Zwanziger:2001kw,Pawlowski:2003hq} and lattice
gauge theory \cite{Bonnet:2000}. The results provide strong support for
these scenarios even though the infrared enhancement of the ghost is a
subject of ongoing debate, for a summary see e.g.\
\cite{Fischer:2008uz,vonSmekal:2008ws}. This paves the way for a
comprehensive understanding of the non-perturbative mechanisms of
strongly-coupled gauge systems.

A pressing open question is the relation of color confinement to quark
confinement. Typical quark-confinement criteria based on the
Wilson-loop or Polyakov-loop expectation value \cite{Polyakov:1978vu}
in quenched QCD have so far remained inaccessible from the pure
knowledge of low-order correlation functions of the gauge sector,
although evidence for a linearly rising potential between static
quarks has been collected within certain approximation schemes, e.g.\
\cite{Zwanziger:1998ez,Alkofer:2006gz,Epple:2006hv}.

{In this letter, we propose a method for computing the full
Polyakov-loop potential from background-field-dependent Green
functions. Our approach relates the order parameter of quark
confinement, the expectation value of the Polyakov loop, to the
momentum dependence of gauge-dependent Green functions. This leads to
a simple confinement criterion in any gauge. The method is explicitly
applied in the Landau gauge, where it relates the KOGZ scenario of
gluon confinement to quark confinement. We evaluate the effective
potential of a purely temporal background field configuration $A_0$,
being directly related to the Polyakov loop variable,}
\begin{equation}
L(x) =\frac{1}{\Nc} \tr\, \CP\, \exp \left( \I g
  \int_0^\beta dx_0\, {A}_0(x_0,x) \right), \label{1}
\end{equation}
where $\CP$ denotes time ordering, and the group trace is taken in the
fundamental representation.  The negative logarithm of the Polyakov
loop expectation value relates to the free energy of a static
fundamental color source. Moreover, $\langle L\rangle$ measures
whether center symmetry is realised by the ensemble under
consideration, see e.g.\ \cite{Svetitsky:1985ye}. A center-symmetric
confining (disordered) ground state ensures $\langle L\rangle=0$,
whereas deconfinement $\langle L\rangle\neq 0$ signals the ordered
phase and center-symmetry breaking.

The order parameter $\langle L[A_0]\rangle $ is conveniently parametrised in
the Polyakov gauge: $\partial_0 {A}_0=0$ with $A_0$ in the Cartan
subalgebra. Then, $\langle{A}_0\rangle$ is sensitive to topological defects
related to confinement \cite{Reinhardt:1997rm}, and also serves as a
deconfinement order parameter. More specifically, $\langle L[A_0]\rangle
  $ is bounded from above by $L[\langle {A_0} \rangle]$ owing to the Jensen
  inequality $L[\langle {A_0} \rangle]= \tr \exp(\I g \beta \langle
  {A}_0\rangle)/\Nc\geq \langle L\rangle$, such that $L[\langle {A_0}
    \rangle]$ is nonzero in the center-broken phase. In the center-symmetric
  phase where the order parameter $\langle L[A_0]\rangle $ vanishes, also the
  observable $L[\langle {A_0} \rangle]$ can be shown to be strictly zero
  \cite{Marhauser:2008fz}. This establishes both $\langle A_0 \rangle$ as well
  as $L[\langle {A_0} \rangle]$ as a deconfinement order parameter. 

In the present work, we compute the effective potential for
$\langle{A}_0\rangle$ from Green functions in the background-field
formalism \cite{Abbott:1981hw} in the Landau-DeWitt gauge by means of
the functional RG. These Green functions can be deduced from that in
the Landau gauge, that is at vanishing background field. Our
construction relates gluon confinement encoded in the IR behaviour of
Green functions to the potential of the order parameter for quark
confinement, and provides a simple confinement criterion.

\section{Background-field flows}

The effective potential is given by $V(L[A_0])=\Gamma/\Omega$, where
$\Gamma$ is the effective action taken at the mean field $A_0$, and
$\Omega$ is the space-time volume.  We evaluate the effective action
$\Gamma$ in the background field approach, where $\Gamma$ on the one
hand depends on the field variable $A$, being the expectation value
of the fluctuating quantum field. On the other hand, a dependence on
an auxiliary background field $\bar A$ is introduced by gauge-fixing
the fluctuating field with respect to the background,
\begin{equation}\label{eq:backgauge}
  D_\mu(\bar A) (A-\bar A)_\mu=0.
\end{equation}
Implementing this gauge condition at vanishing gauge parameter
constitutes the Landau-DeWitt gauge. With the gauge fixing
\eq{eq:backgauge}, the field dependence of the effective action can be
summarized as, $\Gamma=\Gamma[{\Phi},\bar A]$ with fluctuation fields
${\Phi}=(A-\bar A,C,\bar C)$ relative to the background. The
important connection to the standard effective action depending only
on $A$ is established through the identity $\Gamma[A]=
\Gamma[0,\bar A=A]$, \cite{Abbott:1981hw}.

In the present study, we identify the background field with the
Polyakov loop field, $\bar A=A_0$. For evaluating the effective
potential $V(L[A_0])$, it suffices to consider $A_0$ as constant,
yielding
\begin{equation}
V_{k}(L[A_0])= \frac{\Gamma_k[0,A_0]}{\Omega}.
\end{equation}
We compute the effective potential non-perturbatively by means of the
functional RG (FRG) for the effective action \cite{Wetterich:1993yh},
for reviews see \cite{Litim:1998nf,Pawlowski:2005xe}. The flow
equation for $\Gamma[{\Phi},\bar A]$ in the background-field approach
reads
\begin{equation}
  \partial_k \Gk[{\Phi},\bar A]=\frac{1}{2} \Tr \0{1}{\Gamma_k^{(2,0)}
    [{\Phi},\bar A]+R_k} \partial_k R_k\,,
  \label{eq:backflow}
\end{equation}
where $\Gamma_k^{(n,m)}=\0{\delta^{n} }{\delta{\Phi}^n}\0{\delta^{m}}{
  \delta\bar A^m}\Gamma_k$
\cite{Pawlowski:2005xe,Reuter:1993kw,Pawlowski:2001df}. The regulator
function $R_k$ implements an IR regularisation at $p^2\simeq k^2$, and
the trace $\Tr$ sums over momenta, internal indices and species of
fields. The flow \eq{eq:backflow} interpolates between the classical
action in the UV and the quantum effective action
$\Gamma=\Gamma_{k=0}$ in the IR. For ${\Phi}=0$, \Eq{eq:backflow}
entails the flow of $\Gamma_k[A]=\Gamma_k[0,\bar A=A]$, and as a
specifically interesting case, that of $V_{k}(L[A_0])=
\Gamma_k[A_0]/\Omega$.

Background-field flows have been applied successfully to
non-perturbative analyses of chiral properties in full QCD
\cite{Pawlowski:1996ch}, including quantitative estimates of the
critical temperature of the chiral transition from first principles.

The flow \eq{eq:backflow} is solved utilising optimisation ideas
\cite{Litim:2000ci,Pawlowski:2005xe} that minimise the truncation
error. Here, we use a specific optimised regulator
\cite{Pawlowski:2005xe},
\begin{eqnarray}
  R_{\rm opt}=(k^2-
  \Gamma_k^{(2)}[0,\bar A])\theta(k^2-\Gamma_{k=0}^{(2)}[0,\bar A]), 
  \label{eq:optreg}
\end{eqnarray}
supplemented by $k$-dependent fields $\Phi$ such that $
\Gamma_k^{(2)}[0,\bar A]=\Gamma_{k=0}^{(2)}[0,\bar A]$ for
$\Gamma_{k=0}^{(2)}[0,\bar A]>k^2$. With the regulator \eq{eq:optreg}
the flow of the standard effective action
  $\Gamma_k[A]=\Gamma_k[0,\bar A=A]$ is also gauge invariant.

The flow of $\Gamma_k[A]$ can, in principle, be obtained from
  \Eq{eq:backflow} by setting $\Phi=0$ and $\bar A=A$. But, this flow 
is not closed \cite{Pawlowski:2005xe,Pawlowski:2001df}: the right-hand
side of \eq{eq:backflow} depends on $\Gamma_k^{(2,0)}[0,A]\neq
\Gamma_k^{(2)}[A]$, the flow of which cannot be extracted from
$\partial_t \Gamma_k[A]$. This has been neglected in previous
non-perturbative applications \cite{Braun:2005cn} but turns out to be
crucial for confinement.  Hence the key input, the two-point function
$\Gamma_k^{(2,0)}[0,A]$ in the background field, has to be
computed separately.
\begin{figure*}[t]
  \hspace*{0.0cm}
\includegraphics[width=0.41\linewidth]{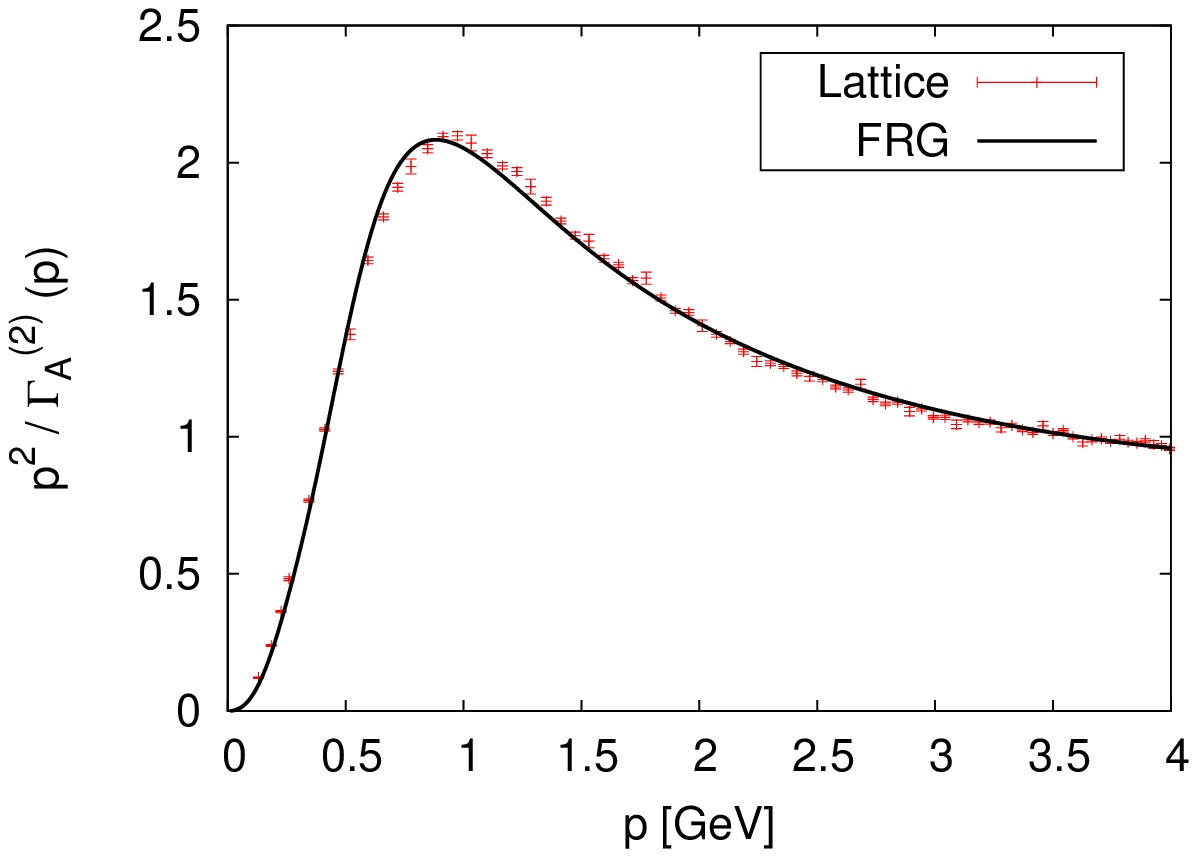}
  \hspace*{2.5cm}
\includegraphics[width=0.41\linewidth]{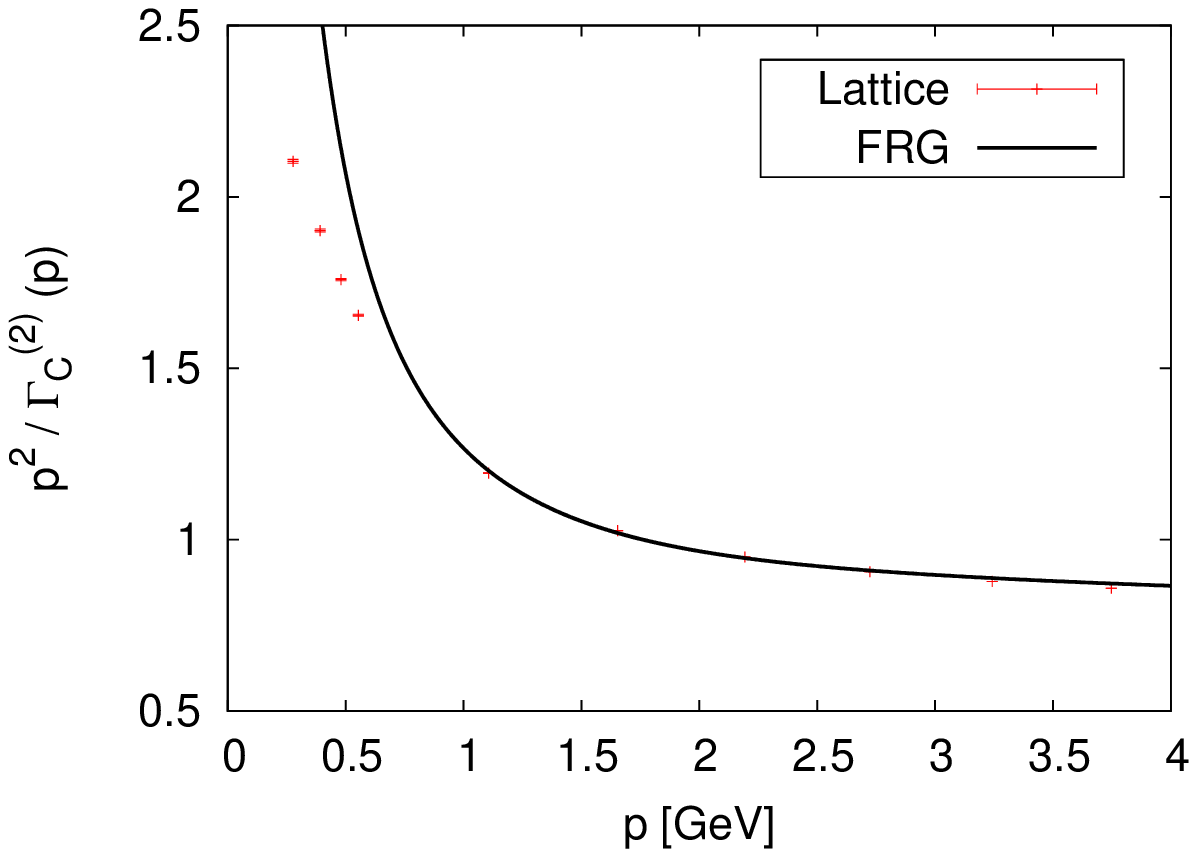} 
\caption{Momentum dependence of the gluon (left panel) and ghost (right panel) 
2-point functions at vanishing temperature. We show the FRG results from 
Ref.~\cite{Fischer:2008uz} (black solid line) and from lattice simulations from 
Ref.~\cite{Bonnet:2000} (red points).}
\label{fig:propagators}
\end{figure*}

\section{Effective action from Landau-gauge propagators} 

First, we observe that in the Landau-DeWitt gauge the longitudinal
components of Green functions decouple from the transversal dynamics,
which further reduces the truncation error, for a detailed discussion
see \cite{Fischer:2008uz}. Moreover, $\Gamma_k^{(2,0)}[0,0](p^2)$
corresponds to the propagator in the Landau gauge, since the
background field gauge with gauge condition \eq{eq:backgauge} reduces
to the Landau gauge for vanishing background field.  The Landau-gauge
propagator has been computed within functional methods,
\cite{vonSmekal1997,Fischer:2002hna,Fischer:2008uz}, as well as
within lattice gauge theory \cite{Bonnet:2000}; for reviews and further
literature, see
\cite{Litim:1998nf,Pawlowski:2005xe,Svetitsky:1985ye%
,Alkofer:2000wg,Fischer:2008uz}.

Recalling the results for Landau-gauge propagators, the gluon
propagator can be displayed as
\begin{eqnarray}\label{eq:gluon} 
\Gamma_{A}^{(2,0)}[0,0](p^2)=p^2 Z_A(p^2) P_{\text{T}}(p) \id +
p^2 \0{Z_{\text L}(p^2)}{\xi} P_{\text{L}} \id ,
\end{eqnarray} 
where $\Pi_{\text{L},\mu\nu}(p)={p_\mu p_\nu}/{p^2}$,
$P_\text{T}=1-P_{\text{L}}$, $\id_{ab}=\delta_{ab}$, and $\xi$ denotes
the gauge parameter. For the ghost, we have
\begin{eqnarray}\label{eq:ghost} 
  \Gamma_{C}^{(2,0)}[0,0](p^2)=p^2 Z_C(p^2)\id.
\end{eqnarray} 
The longitudinal dressing function obeys $Z_{\text L}=1+\CO(\xi)$ and
hence drops out of all diagrams beyond one loop in the Landau
  gauge $\xi=0$. The dressing functions $Z_{A,C}$ encode the
  nontrivial behavior of the full propagators.  In the deep infrared,
  they exhibit the leading momentum behaviour
\begin{eqnarray}\label{eq:ir}
  Z_A(p^2\to 0)\simeq (p^2)^{\kappa_A},\qquad  
  Z_C(p^2\to 0)\simeq (p^2)^{\kappa_C}. 
\end{eqnarray} 
In the last years it has become clear that Landau gauge Yang-Mills
admits a one-parameter family of infrared solutions consistent with
renormalisation group invariance \cite{Fischer:2008uz}. Despite some
formal progress the full understanding of the underlying structure is
a subject of current research. Technically, the parameter corresponds
to an infrared boundary condition, the value of $Z_C(0)$, and is also
relates to $Z_A(p^2\to 0)$ \cite{Fischer:2008uz}. This fact is
reflected in recent lattice solutions \cite{Maas:2009se} and
indications thereof have also been seen in the strong coupling limit
\cite{Sternbeck:2008mv}. 
For $Z_C(p^2 \to 0)\to 0$ it can be shown that there is a unique
scaling solution, \cite{Fischer:2006vf,Alkofer:2008jy}. Then the two
exponents are related and obey the sum rule
\begin{equation}
  0=\kappa_A + 2\kappa_C +\frac{4-d}{2},\label{eq:sumrule}
\end{equation}
in $d$ dimensional spacetime
\cite{Lerche:2002ep,Zwanziger:2001kw,Fischer:2006vf}.  Possible
solutions are bound to lie in the range $\kappa_C\in [1/2\,,\,1]$, see
\cite{Lerche:2002ep}. For the truncation used in most DSE and FRG
computation, we are led to 
\begin{equation}\label{eq:scaling} 
\kappa_C=0.595... \, \quad {\rm and} \quad \kappa_A=- 1.19...\,,
\end{equation}
being the value for the optimised regulator \cite{Pawlowski:2003hq}.
The regulator dependence in FRG computations leads to a range of
$\kappa_C\in [0.539\,,\, 0.595]$, see \cite{Pawlowski:2003hq}; for a
specific flow, see \cite{Fischer:2004uk}. These results entail the
KOGZ confinement scenario: the gluon is infrared screened, whereas the
ghost is infrared enhanced with $\kappa_C>1/2$.

{In turn it can be shown that for non-vanishing $Z_C(0)$ the gluon
propagator tends to a constant in the infrared, $ p^2 Z_A(p^2) \to
m^2$, for related work see e.g.\
\cite{Fischer:2008uz,Cornwall:1981zr,Zwanziger:2009je,Dudal:2008sp,%
  Huber:2009tx,Kondo:2009ug,Boucaud:2006if}. Note that the gluon
propagator then does not correspond to the propagator of a massive
physical particle. Instead, we observe clear indications for
positivity violation in the numerical solutions for the gluon
propagator related to gluon confinement,
\cite{Fischer:2008uz,Cucchieri:2004mf}. Still the gluon decouples from
the dynamics as does a massive particle, hence the name decoupling
solution. The value of $Z_C$ seems to be bounded by its perturbative
value from above, and the gluon mass parameter is bounded from below
\cite{Fischer:2008uz}. The qualitative infrared behaviour is then
given by the infrared exponents}
\begin{equation}
  \kappa_A=-1\,, \quad {\rm and} \quad 
  \kappa_C=0\,.\label{eq:decoupling}
\end{equation}
{We emphasise that even though the infrared exponents for the scaling
solution \eq{eq:scaling} and the decoupling solution
\eq{eq:decoupling} are rather different, the propagators do only
differ in the deep infrared.  It has been suggested in
\cite{Fischer:2008uz} that the infrared boundary condition is directly
related to the global part of the gauge fixing, and hence to different
resolutions of the Gribov problem.  Indeed in \cite{Maas:2009se} the
infrared boundary condition has been implemented directly as a global
completion of the gauge fixing. Note also, that for Landau gauge
Yang-Mills with standard local BRST invariance the requirement of
global BRST singles out the scaling solution. The existence of such a
formulation on the lattice has been shown recently in
\cite{vonSmekal:2007ns}. In summary the results are affirmative for
the above interpretation and are supported by results in the
strong-coupling limit \cite{Sternbeck:2008mv} for different
implementations of lattice Landau gauge.}

{In turn, it has been also shown in a series of works that an infrared
condition also is present in Landau gauge Yang-Mills with the horizon
function, e.g.\
\cite{Zwanziger:2009je,Dudal:2008sp,Huber:2009tx,Kondo:2009ug}. The
latter introduces an explicit (or soft) breaking of BRST invariance as
it restricts the functional integral to the first Gribov region. Still
this does not fix global gauge degrees of freedom as also the first
Gribov region contains infinite many gauge copies. The possibility of
a scaling solution in this framework hints at the validity of
Zwanziger proposal: full BRST invariance is recovered in the
thermodynamic limit if the path integral is restricted to the
fundamental modular domain with only one gauge copy.}

In summary a consistent picture has emerged with nicely relates all
current results. The confirmation of this picture certainly would
provide further insight to the confinement mechanism. For the present
work, we simply note that the scaling solution is singled out by
global BRST invariance which allows the construction of a physical
Hilbert space from gauge fixed correlation functions. Nonetheless, the
whole one-parameter family provides consistent gauge-fixed correlation
functions of Yang-Mills theory and physical observables should be
insensitive to the parameter choice. In the present work, we can test
this statement.

We proceed by extending the Landau-gauge propagator to that in a given
background $\bar A$. The Landau-gauge two-point function
$\Gamma_k^{(2,0)}[0,0](p^2)$ is, apart from its Lorentz structure
provided by the projection operators $P_{\rm T/L}(p)$, a function of
only the momentum squared $p^2$, cf. \Eq{eq:gluon}. At vanishing
temperature, the background field propagator $\Gamma_k^{(2,0)}[0,A]$
can be related to the Landau-gauge propagator in a unique fashion
owing to gauge covariance,
\begin{equation}\label{eq:genprop}
(\Gamma_{k}^{(2,0)}[0,A])^{ab}_{\mu\nu}=(\Gamma_{k}^{(2,0)}[0,0]
(-D^2) )^{ab}_{\mu\nu}+
F^{cd}_{\rho\sigma} f_{\mu\nu\rho\sigma}^{abcd}(D),
\end{equation}
with non-singular $f(0)$ in order to ensure the proper limit of a
vanishing background. The projection operators $P_{\rm T/L}$
implicitly contained in $\Gamma_{k}^{(2,0)}[0,0] (-D^2)$ generalize to
projectors on transversal and longitudinal spaces respectively with
respect to the covariant momentum $D$, $P_{\rm T/L}=P_{\rm T/L}(D)$.
The $f$ terms cannot be obtained from the Landau-gauge propagator, but
are related to higher Green functions {in Landau-DeWitt gauge}.
However, fortunately they do not play a r$\hat {\rm o}$le for our
purpose.

At finite temperature, the Polyakov loop $L$ is a further invariant,
and the $00$ component of the gluon two-point function \eq{eq:genprop}
receives further contributions proportional to derivatives of $L$. For
constant fields $A_0$, we arrive at
\begin{equation}\label{eq:genpropA0}
(\Gamma_{k}^{(2,0)}[0,A_0])^{ab}_{\mu\nu}=(\Gamma_{k}^{(2,0)}[0,0]
(-D^2) )^{ab}_{\mu\nu}+
L\text{-terms}
\,, 
\end{equation} 
as the $f$ term in \eq{eq:genprop} vanishes: $F(A_0)=0$. In this
letter, we take only the explicit $T$ dependence due to Matsubara
frequencies into account and drop any implicit $T$ dependence: first,
this amounts to dropping the $L$ contribution in \eq{eq:genpropA0}.
This term is related to the second derivative of the effective
potential $V_k^{(2)}$ via Nielsen identities
\cite{Pawlowski:2005xe,Pawlowski:2001df}, and can indeed be estimated
by $V_k^{(2)}$. Its influence on the confinement-deconfinement phase
transition temperature is parametrically suppressed, and can be
neglected for a first estimate of the critical temperature
$T_{\text{c}}$. Second, this amounts to using the zero-temperature
propagators. First results indeed indicate that transversal and
longitudinal gluon and ghost propagators are little modified
\cite{Gruter:2004bb,Cucchieri:2007ta,Maas:2009fg} for higher Matsubara
frequencies $2 \pi T n$ for $n>2,3$. The biggest change appears in the
gluon propagator longitudinal with respect to the heat bath that
develops some enhancement compared to the transversal counterpart. The
inclusion of the full temperature dependence is necessary for an
accurate determination of, e.g., the critical exponents or the
equation of state (see, e.g., \cite{Chernodub:2007rn}). This will be
subject of a forthcoming paper.
\begin{figure*}[t]
  \hspace*{0.0cm}
\includegraphics[width=0.41\linewidth]{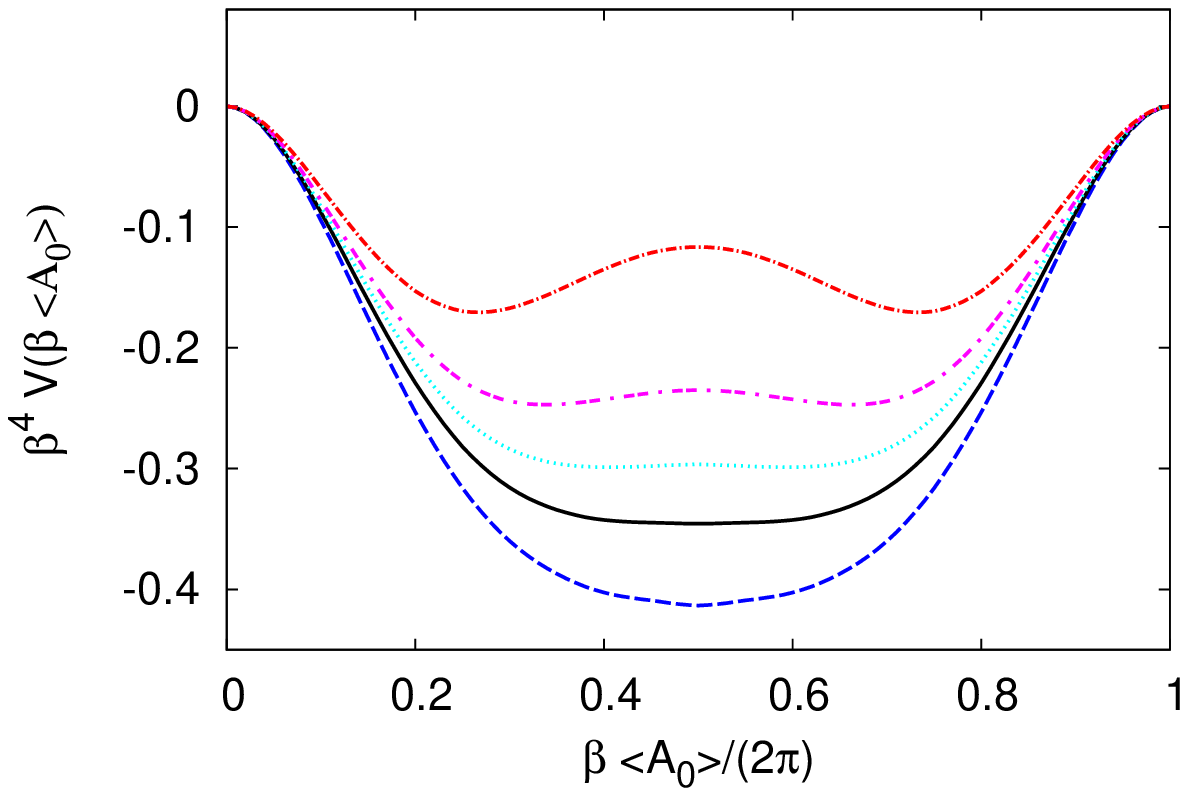}
  \hspace*{2.5cm}
\includegraphics[width=0.41\linewidth]{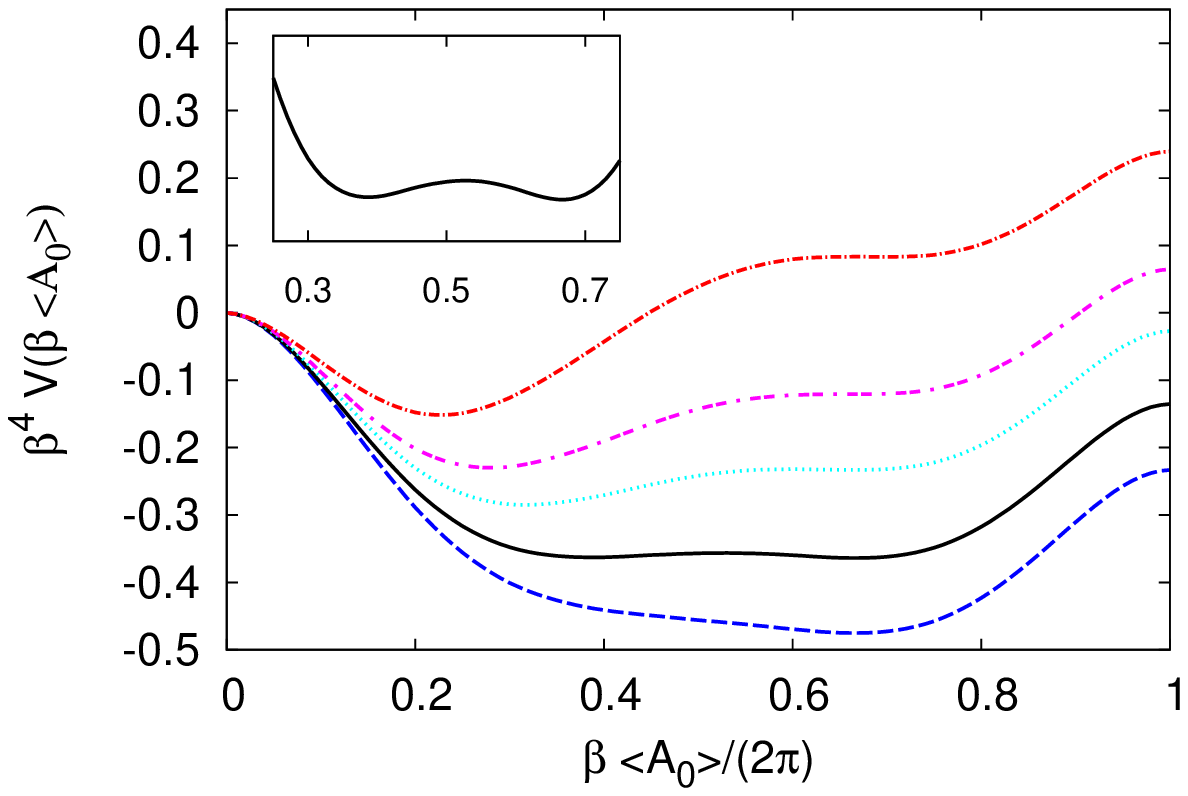} 
\caption{ Order-parameter potential for SU(2) (left panel) and SU(3)
  (right panel) for various temperatures. For SU(2) we show the
  potential for $T=260,\, 266,\, 270,\, 275,\, 285\,\text{MeV}$ (from
  bottom to top). We find $T_c\approx 266\,\text{MeV}$ for SU(2). In
  case of SU(3), the relevant minima occur in the $A_0^8$ direction in
  the Cartan subalgebra. A slice of the potential in this direction is
  shown for $T=285,\, 289.5,\, 295,\, 300,\, 310\,\text{MeV}$ (from
  bottom to top). A magnified view on the potential at the phase
  transition is shown in the inlay, revealing the 1st-order nature of
  the phase transition with two equivalent minima at at $T_c\approx
  289.5\,\text{MeV}$.}
\label{fig:potentials}
\end{figure*}

\section{A simple order-disorder confinement criterion}

The preceding analysis gives rise to a simple confinement criterion which
relates the IR behaviour of gluon and ghost 2-point functions to the
deconfinement order parameter. Integrating the flow \eq{eq:backflow}, we
obtain
\begin{equation}
\Gamma[A]=\frac{1}{2} \Tr \ln \Gamma^{(2,0)}[0,A] + \mathcal{O}(\pat
\Gamma_k^{(2,0)}) + \text{c.t}.,\label{eq:flowint}
\end{equation}
where the counterterms (c.t.) denote the appropriate UV initial
conditions of the flow, and the $O(\pat \Gamma_k^{(2,0)})$ terms
correspond to integrated RG improvement terms. The first term is
explicitly regulator-independent, and so is the improvement term. This
can be used to show within the specific choice \eq{eq:optreg} that the
improvement term is subdominant for the following analytic argument,
which is confirmed by the full numerical solution.

The effective action in \eq{eq:flowint} involves the Lapacian $-D^2 $
for vanishing field strength. In the constant ${A}_0$ background, we
use the parametrisation $g {A}_0^a= 2\pi T \phi^a$, where $\phi^a$ is
a vector in the Cartan subalgebra.  The spectrum of the Laplacian then
reads
\begin{equation}
\text{spec}\{-D^2[{A}_0]\} =\vec{p}^2 + (2\pi T)^2 (n -
\nul |\phi|)^2,\label{spec}
\end{equation}
where the $\nul$ denote the $\Nc^2-1$ eigenvalues of the hermitian
color matrix $T^a \phi^a/|\phi|$, $(T^a)^{bc}=-\I f^{abc}$ being the
generators of the adjoint representation. From \Eqref{spec}, it is
clear that $\phi$ is a compact variable.

At high temperature, $2 \pi T\gg \Lambda_{\rm QCD}$, the effective potential
is dominated by the perturbative regime, and the background-covariant inverse
propagators of both gluons and ghosts are approximately given by
their tree-level values $\Gamma^{(2),\text{tree}}(-D^2)=-D^2 $. The
perturbative limit of the effective potential $V$ in $d>2$ is given by the
well-known Weiss potential \cite{Weiss:1980rj},
\begin{eqnarray}
V^{\text{UV}}(\phi^a)&=& \left\{ \frac{d-1}{2} + \frac{1}{2} - 1
\right\} \frac{1}{\Omega} \Tr\ln \big(-D^2[{A}_0]\big)
 \label{eq:Weiss}\\
&=& -\frac{(d-2)\Gamma(d/2)}{\pi^{d/2}}\, T^d
\sum_{l=1}^{\Nc^2-1} \sum_{n=1}^\infty \frac{\cos 2\pi n \nul |\phi|
}{n^d},\nonumber
\end{eqnarray}
where the terms in curly brackets in the first row denote the
contributions from transversal gluons, longitudinal gluons and ghosts,
respectively. In the second row, we have dropped a $T$- and
field-independent constant.  The Weiss potential exhibits maxima at
the center-symmetric points where $L[\langle A_0\rangle]=0$, implying
that the perturbative ground state is not confining, $\langle L
\rangle \neq0$. 

Now, we perform the same analysis at low temperature $2\pi T\ll
\Lambda_{\rm QCD}$. The series in \eq{eq:Weiss} converges rather
rapidly due to the $1/n^d$ suppression of higher terms. Hence, the
effective potential $V(\phi^a)$ is dominantly induced by fluctuations
with momenta near the temperature scale $p^2\sim (2\pi
T)^2$. {This does not change qualitatively in the presence of a
  non-trivial momentum dependence of the propagators. We conclude that
  only the first 10-20 Matsubara frequencies play a r$\hat{\rm
    o}$le. Moreover, changing the propagator for the first two or three
  Matsubara frequencies, even though their weight is higher, only
  gives rise to minimal changes in the potential. This fully justifies
  the zero-temperature estimate on the propagators.}

With the parametrisation \eq{eq:gluon},\eq{eq:ghost}, the dressing
functions $Z_A(p^2),Z_C(p^2)$ in the KOGZ scenario are characterised
by the power-law behaviour \eq{eq:ir} in the deep IR, $p^2\ll
\lqcd^2$. For low enough temperature, the spectral window
$-D^2[A_0]\simeq (2 \pi T)^2$ is in this asymptotic regime, and thus
the effective potential arises dominantly from fluctuations in the
deep IR,
\begin{eqnarray}
V^{\text{IR}}(\phi^a)&=& \left\{ \frac{d-1}{2} (1+\kappa_A) +
  \frac{1}{2} -(1+\kappa_C)\right\}  \nonumber\\
 && \quad \times \frac{1}{\Omega} \Tr\ln
  \big(-D^2[{A}_0]\big)  \label{eq:WeissIR}
\end{eqnarray}
\begin{eqnarray}
&=& \left\{ 1+ \frac{(d-1)\kappa_A-2\kappa_C }{d-2} \right\}
V^{\text{UV}}(\phi^a).\nonumber
\end{eqnarray}
If the anomalous dimensions are such that the expression in curly
brackets becomes negative, the effective potential is reversed and the
confining center-symmetric points become order-parameter minima. 
 
We conclude that the effective action \eqref{eq:WeissIR} predicts a
center-symmetric quark-confining ground state if 
\begin{equation}\label{eq:confcrit}
f(\kappa_A,\kappa_c;d)=d-2 + (d-1)\kappa_A-2\kappa_C<0.
\end{equation}
Provided that the $\mathcal O (\pat\Gamma_k(2,0))$ terms in
\Eqref{eq:flowint} remain subdominant, this equation provides a
simple, necessary and sufficient criterion for quark confinement in
Yang-Mills theory: if \Eqref{eq:confcrit} is satisfied the order
parameter for quark confinement vanishes, $\langle
L[A_0]\rangle=0$. It is satisfied for the whole one-parameter family
of infrared solutions of Landau-gauge Yang-Mills theory. For the
scaling solution with the sum rule \eq{eq:sumrule}, we are led to
\begin{equation}
\kappa\equiv \kappa_C > \frac{d-3}{4}.
\label{eq:critscaling}
\end{equation}
which is satisfied for the numerical values for the scaling
exponents $\kappa_d$ in $d=2,3,4$, see 
\cite{Lerche:2002ep,Zwanziger:2001kw}. Specifically in $d=4$, we have
\Eqref{eq:scaling}, and hence 
\begin{equation}
f(-2 \kappa_c,\kappa_c;4)=-2.76...\,.
\label{eq:fcaling}
\end{equation}
For the decoupling solution \eq{eq:decoupling}, we are led to
\begin{equation}
f(-1,0;d)=-1\,.
\label{eq:fdecoupling}
\end{equation}
Both values imply confinement, and hence the whole one parameter
family of solutions is confining. Note that this is to be expected as
corresponding propagators can be obtained within lattice simulations
with different gauge fixings.

The above confinement criterion has to be compared to the Kugo-Ojima
criterion for color confinement $\kappa>0$ and the Zwanziger horizon
condition for the ghost $\kappa>0$ and for the gluon $\kappa>1/2$ in
$d=4$. The Kugo-Ojima criterion and the Zwanziger horizon condition
are necessary but not sufficient for confinement. Indeed for
${0<}\kappa<1/4$ in four dimensions, we observe that the Kugo-Ojima
criterion is satisfied but does not lead to confinement according to
the present confinement criterion \eq{eq:critscaling}. We would also
like to emphasise that, in effective theories for QCD,
\Eqref{eq:confcrit} only serves as a necessary condition. It only
restricts the propagators, and other Green functions in effective
theories might violate related constraints.

\section{Results for the phase transition}

In contradistinction to the simple confinement criterion put forward
above, the physics of the confinement-deconfinement phase transition,
e.g., the transition temperature and the order of the phase
transition, is determined by the dynamics of the system and not by its
IR asymptotics.  Indeed, we find that fluctuations in the
non-perturbative mid-momentum regime induce the center-symmetric
minimum of the ${A_0}$ potential long before the propagators acquire
their deep IR scaling form \eqref{eq:ir}. As only the deep infrared is
sensitive to the infrared boundary condition the critical temperature
is insensitive to this choice which is confirmed in the explicit
computation.

The results presented below are achieved by numerically integrating
the flow equation \eqref{eq:backflow} in order to obtain the potential
for an ${A_0}$ background. The present truncation is optimised by
using Landau-gauge propagators and RG improvement terms at zero
temperature computed from the FRG for different infrared boundary
conditions. It is also compared to results obtained by using fits to
Landau-gauge propagators as measured by lattice gauge theory
\cite{sternbeck06} and the RG improvement computed in
\cite{Fischer:2008uz}. For our numerical study of the order-parameter
potential we have suitably amended the lattice propagators by the
perturbative behaviour in the UV and the corresponding power laws
\eqref{eq:ir} in the IR. In Fig.~\ref{fig:propagators} we show the
gluon and ghost propagators as obtained from FRG
computations~\cite{Fischer:2008uz} and lattice
simulations~\cite{sternbeck06}. {There is an impressive agreement
  of the results for the ghost and gluon propagators for momenta
  larger than about $p\gtrsim 700\,\text{MeV}$ which holds for the
  whole one parameter family of solutions including the scaling
  one. The results for the ghost dressing from scaling solution of the
  FRG and lattice simulations start deviating for {$p\lesssim
    700\,\text{MeV}$} whereas the scaling solution for the gluon
  starts deviating for even lower momenta. Since the lowest
  non-vanishing Matsubara mode is associated with momenta at about
  $|p|\sim 2\pi T_{\text{c}} \sim 1700\,\text{MeV}$, the differences
  in the IR are hardly probed in the present study of the
  deconfinement phase transition. This is confirmed by the explicit
  computation. In the vacuum limit, $T\to0$, the picture arising from
  the preceding simple confinement criterion is confirmed: a
  sufficient amount of gluon screening with or without an IR
  enhancement of the ghost creates a center-disordered ground state
  with quark confinement.}

\begin{figure}[t]
\includegraphics[width=1\linewidth]{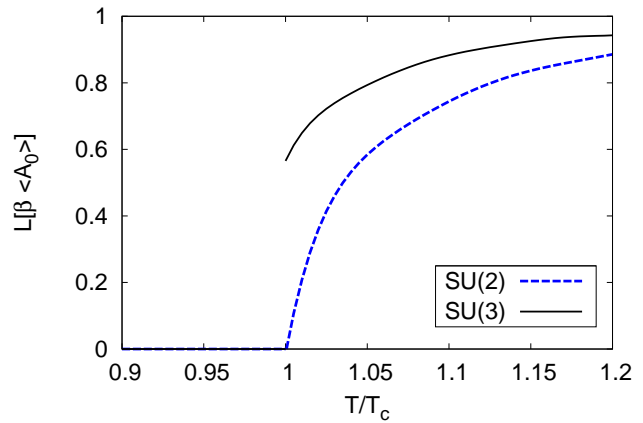}
\caption{Polyakov loop for the ${A_0}$ expectation value
  $L[\beta\langle {A_0}\rangle]$ for SU(2) (blue/dashed line) and
  SU(3) (black/solid line). The phase transition is of second order
  for SU(2) and of first order for SU(3).}
\label{fig:su3}
\end{figure}

The confinement-deconfinement transition is taking place in the
mid-momentum regime that interpolates between the perturbative regime
and the IR asymptotics. The effective potentials for SU(2) and SU(3)
for various temperature values near the phase transition are displayed
in Fig.~\ref{fig:potentials}. For SU(3) (right panel), the slice of
the potential in $A_0^8$ direction is depicted where the relevant
minima for the phase transition occur. Reading off $\langle {A_0}
\rangle$ from the minimum of the potential at a given temperature, we
can determine $L [\langle {A_0} \rangle]$ which is plotted in
Fig.~\ref{fig:su3}.  For SU(2) (blue/dashed line), the phase
transition is of second order.  For SU(3) (black/solid line), we
clearly observe a first-order phase transition at a critical
temperature of $T_{\text{c}}\simeq 284\pm 10$MeV with a lattice string
tension $\sqrt{\sigma}=440$MeV, that is
$T_{\text{c}}/\sqrt{\sigma}=0.646\pm 0.023$. The error relates to the
uncertainties of the fits for the lattice propagators which exceed the
estimate on the systematic error in the FRG computation. The result
compares favourably both qualitatively and quantitatively with lattice
simulations, see e.g.\ \cite{Fingberg:1992ju,sternbeck06}. Also, our
result for $L[\langle {A_0} \rangle]$ in the deconfined phase is
higher than the lattice measurement of the Polyakov-loop expectation
value $\langle L \rangle$ in agreement with the Jensen inequality $L
[\langle {A_0} \rangle] > \langle L \rangle$. Note however that this
statement has to be taken with care as the lattice result involves a
non-trivial renormalisation factor which is absent in the definition
of $L [\langle {A_0} \rangle]$. Indeed, $L [\langle {A_0} \rangle]\leq
1$ whereas the renormalised Polyakov loop $\langle L \rangle_{\rm
  ren}$ necessarily exceeds unity for some temperature range as can be
deduced from perturbation theory.

As discussed above, corrections to our estimate arise from finite-$T$
modifications of the propagators as well as from order-parameter
fluctuations; the latter are more pronounced for SU(2) owing to the
second-order nature of the transition. As expected, the critical
temperature is not sensitive to the one-parameter family of solutions,
it is only sensitive to the mid-momentum regime at about 1
GeV. Indeed, this also explains the fact that the gluon mass parameter
is restricted from below: small gluon mass parameters would also
trigger changes in the mid-momentum regime and almost certainly change
physical quantities such as the critical temperature.

In summary, we have established a simple confinement criterion that
relates quark confinement to the infrared behaviour of ghost and gluon
Green functions.  This confinement criterion is applicable in
arbitrary gauges. Our full numerical analysis of the IR dynamics
predicts a second-order phase transition for SU(2) and a first-order
phase transition for SU(3), the critical temperature of which is in
quantitative agreement with lattice results. The related Polyakov loop
potential also plays an important r$\hat o$le for full QCD
computations with dynamical quarks within functional methods, for
first results on the QCD phase diagram see \cite{Braun:2009gm}.

{\em Acknowledgements --} We thank K.~Langfeld, A.~Sternbeck, L.~von
Smekal and I.-O.~Stamatescu for providing lattice data and useful
discussions. HG acknowledges DFG support under Gi 328/1-4.  JB
acknowledges support by the Natural Sciences and Engineering Research
Council of Canada (NSERC). TRIUMF receives federal funding via a
contribution agreement through the National Research Council of
Canada.

\end{document}